# Estimating strength of DDoS attack using various regression models

## B.B. Gupta*, R.C. Joshi and Manoj Misra

Department of Electronics and Computer Engineering,
Indian Institute of Technology Roorkee,
Roorkee 247667, India
E-mail: gupta.brij@gmail.com
E-mail: rcjosfec@iitr.ernet.in
E-mail: manojfec@iitr.ernet.in
*Corresponding author

**Abstract:** Anomaly-based DDoS detection systems construct profile of the traffic normally seen in the network, and identify anomalies whenever traffic deviate from normal profile beyond a threshold. This extend of deviation is normally not utilised. This paper reports the evaluation results of proposed approach that utilises this extend of deviation from detection threshold to estimate strength of DDoS attack using various regression models. A relationship is established between number of zombies and observed deviation in sample entropy. Various statistical performance measures, such as coefficient of determination ($R^2$), coefficient of correlation (CC), sum of square error (SSE), mean square error (MSE), root mean square error (RMSE), normalised mean square error (NMSE), Nash-Sutcliffe efficiency index ($\eta$) and mean absolute error (MAE) are used to measure the performance of various regression models. Internet type topologies used for simulation are generated using transit-stub model of GT-ITM topology generator. NS-2 network simulator on Linux platform is used as simulation test bed for launching DDoS attacks with varied attack strength. A comparative study is performed using different regression models for estimating strength of DDoS attack. The simulation results are promising as we are able to estimate strength of DDoS attack efficiently with very less error rate using various regression models.

**Keywords:** DDoS attack; intrusion detection; regression analysis; entropy.

**Biographical notes:** B.B. Gupta received his Bachelor's degree in Information Technology in 2005 from Rajasthan University, India. He is currently a PhD student in the Department of Electronics and Computer Engineering at Indian Institute of Technology, Roorkee, India. He has been awarded Canadian Commonwealth Scholarship (CCSP) and Government of Canada Awards (GCA) in 2009. His research interests include defence mechanisms for thwarting denial of service attacks, network security, cryptography, data mining and data structure and algorithms.

R.C. Joshi received his Bachelor's degree in Electrical Engineering from Allahabad University, India in 1967. He received his Master's and PhD in Electronics and Computer Engineering from University of Roorkee, India in 1970 and 1980, respectively. Currently, he is working as a Professor at the Indian Institute of Technology Roorkee, India. He has a vast teaching experience exceeding 38 years at graduate and postgraduate levels at IIT Roorkee. Presently, he is actively involved in research in the fields of database management system, data mining, bioinformatics, information security and mobile computing.

Manoj Misra received his Bachelor's degree in Electrical Engineering in 1983 from HBTI Kanpur, India. He received his Master's and PhD degree in Computer Engineering in 1986 and 1997 from University of Roorkee, India and Newcastle upon Tyne, UK, respectively. He is currently a Professor at Indian Institute of Technology Roorkee. He has guided several PhD theses, ME/MTech dissertations and completed various projects. His areas of interest include mobile computing, distributed computing and performance evaluation.

## 1 Introduction

Today, DoS attacks and more particularly the distributed ones (DDoS) are one of the latest threat and pose a grave danger to users, organisations and infrastructures of the internet. In these attacks, goal of the attacker is to tie up chosen key resources at the victim, usually by sending a high volume of seemingly legitimate traffic requesting some services from the victim (Gupta et al., 2008). The first publicly reported DDoS attack appeared in the late 1999 against a university. These attacks quickly became increasingly popular as communities of crackers developed and released automated tools to carry them out. This made attack process within inexperienced crackers' capabilities. Thus, these attacks are easiest to implement from an attacker's point of view and definitely one of the costliest from a business point of view. One of the major challenges in defending against DDoS attacks is to accurately detect their occurrences in the first place. Anomaly-based DDoS detection systems construct profile of the traffic normally seen in the network, and identify anomalies whenever traffic deviate from normal profile beyond a threshold (Gupta et al., 2009). This extend of deviation is normally not utilised. Proposed approach utilises this extend of deviation from detection threshold, to estimate strength of DDoS attack using various regression models, i.e., linear, exponential, power, logarithmic and polynomial. A real time estimation of the strength of DDoS attack is helpful to suppress the effect of attack by choosing predicted strength of most suspicious attack sources for either filtering or rate limiting. Moore et al. (2006) have already made a similar kind of attempt, in which they have used backscatter analysis to estimate number of spoofed addresses involved in DDoS attack. This is an offline analysis based on unsolicited responses.

Our objective is to find the relationship between strength of DDoS attack and deviation in sample entropy. In order to estimate strength of DDoS attack, several models are developed using various regression techniques. For each regression model, we have calculated various statistical performance measures, i.e., $R^2$, CC, SSE, MSE, RMSE, NMSE, $\eta$, MAE and residual error. A comparative study is performed using different regression models for estimating strength of DDoS attack. Internet type topologies used for simulation are generated using transit-stub model of GT-ITM topology generator. NS-2 network simulator (NS Documentation, 2010) on Linux platform is used as simulation test bed for launching DDoS attacks with varied attack strengths. In our simulation experiments, total number of zombies is fixed to 100 and attack traffic rate range between 10 Mbps and 100 Mbps; therefore, mean attack rate per zombie is varied from 0.1 Mbps to 1 Mbps.

The remainder of the paper is organised as follows: Section 2 contains overview of various regression models. Section 3 presents various statistical performance measures. Intended detection scheme are described in Section 4. Section 5 describes experimental setup and performance analysis in details. Model development is presented in Section 6. Section 7 contains simulation results and discussion. Finally, Section 8 concludes the paper.

## 2 Regression models

Regression analysis (Lindley, 1987; Freedman, 2005) is a statistical tool for the investigation of relationships between variables. Usually, the investigator seeks to ascertain the causal effect of one variable upon another. More specifically, regression analysis helps us to understand how the typical value of the dependent variable changes when any one of the independent variables is varied, while the other independent variables are held constant. Variables which are used to 'explain, other variables are called explanatory variables. Variable which are explained are called response variable. A response variable is also called a dependent variable, and an explanatory variable is sometime called an independent variable, or a predictor, or repressor. When there is only one explanatory variable the regression model is called a simple regression, whereas if there are more than one explanatory variable the regression model is called multiple regression.

### 2.1 Different types of regression model used

### 2.1.1 Linear regression

Linear regression (Montgomery et al., 2001; Jammalamadaka, 2003) includes any approach to modelling the relationship between a dependent variable $Y$ and one or more independent variables denoted $X$, such that the model depends linearly on the unknown parameters to be estimated from the data. Such a model is called a linear model. Linear regression uses one independent variable to explain and/or predict the outcome of $Y$, while multiple regression uses two or more independent variables to predict the outcome. The general form of linear of regression is:

$$M1: Y_i = \hat{Y}_i + \varepsilon_i$$
$$\hat{Y}_i = \beta_0 + \beta_1 X_i, \quad (1)$$

where

$Y$  is dependent variable

$X$  is independent/predictor variable

$\beta_0$  is intercept

$\beta_1$  is slope

$\varepsilon$ is regression residual.

To estimate the intercept and slope, minimise sum of square error (SSE)

$$SSE = \sum \varepsilon_i^2 = \sum (Y_i - \hat{Y}_i)^2 = \sum (Y_i - \beta_0 - \beta_1 X_i)^2$$

$$\frac{\partial SSE}{\partial \beta_0} = \frac{\sum 9Y_i - \beta_0 - \beta_1 X_i^2}{\partial \beta_0} = -2 \sum (Y_i - \beta_0 - \beta_1 X_i) = 0$$

$$\Longrightarrow \hat{\beta}_0 = \bar{Y} - \hat{\beta}_1 \bar{X}$$

$$\frac{\partial SSE}{\partial \beta_1} = \frac{\sum (Y_i - \bar{Y} + \beta_1 \bar{X} - \beta_1 X_i)^2}{\partial \beta_1} = -2 \sum (X_i - \bar{X})(Y_i - \bar{Y} + \beta_1 \bar{X} - \beta_1 X_i) = 0$$

$$\Longrightarrow \hat{\beta}_1 = \frac{\sum (X_i - \bar{X})(Y_i - \bar{Y})}{\sum (X_i - \bar{X})^2}$$

### 2.1.2 Polynomial regression

Polynomial regression (Stigler, 1971; Anderson, 1962) is a form of regression in which the relationship between the independent variable *X* and the dependent variable *Y* is modelled as an *i*th order polynomial. The general form of this regression model is as follows:

$$M2 : Y_i = \hat{Y}_i + \varepsilon_i$$
$$\hat{Y}_i = \beta_0 + \beta_1 X + \beta_2 X^2 + \ldots\ldots + \beta_n X^n \tag{2}$$

### 2.1.3 Logarithmic regression

A logarithmic regression (Baskerville, 1972; Aneuryn-Evans and Deaton, 1980; El-Shaarawi and Viveros-Aguilera, 2006) also known as logarithmic least squares fittings. For the relation between dependent and independent variables, it finds the logarithmic function that best fits a given set of data points. A logarithmic equation has the following general form:

$$M3 : Y_i = \hat{Y}_i + \varepsilon_i$$
$$\hat{Y}_i = \beta_0 Ln(X_i) + \beta_1 \tag{3}$$

Logarithmic data will exhibit a straight-line relationship when graphed with the *X* values on a log scale and the *Y* values on a linear scale.

### 2.1.4 Power regression

Power regression (Gowariker et al., 1991), also known as log-log regression, takes the input signal and fits the function

$$M4 : Y_i = \hat{Y}_i + \varepsilon_i$$
$$\hat{Y}_i = \beta_0 . X_i^{\beta_1} \qquad (4)$$

to it where $X$ is the variable along the x-axis.

The function is based on the linear regression, with both axes scaled logarithmically. Power regressions will not allow an independent variable value of zero.

### 2.1.5 Exponential regression

An exponential regression (Melas, 1978; Dette et al., 2006), also known as exponential least squares fittings. For the relation between two variables, it finds the exponential function that best fits a given set of data points. Exponential regression takes the input signal and fits an exponential function

$$M5 : Y_i = \hat{Y}_i + \varepsilon_i$$
$$\hat{Y}_i = \beta_0 . e^{\beta_1 X_i} \qquad (5)$$

to it where $X$ is the variable along the x-axis.

### 2.2 Input and output

For regression models, a relationship is developed between strength of DDoS attack $Y$ (output) and observed deviation in sample entropy $X$ (input). Here, $X$ is equal to ($H_c - H_n$). Our proposed regression-based approach utilises this deviation in sample entropy $X$ to estimate strength of DDoS attack using various regression models, i.e., linear, exponential, power, logarithmic and polynomial.

## 3 Statistical performance measures

The different statistical parameters of each model are adjusted during calibration to get the best statistical agreement between observed and simulated variables. For this purpose, various performance measures, such as coefficient of determination ($R^2$), coefficient of correlation (CC), sum of square error (SSE), mean square error (MSE), root mean square error (RMSE), normalised mean square error (NMSE), Nash-Sutcliffe efficiency index ($\eta$) and mean absolute error (MAE) are used to measure the performance of various models. These measures are defined below:

1  *$R^2$:* $R^2$ is a descriptive measure of the strength of the regression relationship, a measure how well the regression line fit to the data. $R^2$ is the proportion of variance in dependent variable which can be predicted from independent variable.

$$R^2 = \frac{\left(\sum_{i=1}^{N}(Y_o - \bar{Y}_o)(Y_c - \bar{Y}_c)\right)^2}{\left[\sum_{i=1}^{N}(Y_o - \bar{Y}_o)^2 \cdot \sum_{i=1}^{N}(Y_c - \bar{Y}_c)^2\right]} \qquad (6)$$

2  *CC:* The CC can be defined as:

$$CC = \frac{\sum_{i=1}^{N}(Y_o - \bar{Y}_o)(Y_c - \bar{Y}_c)}{\left[\sum_{i=1}^{N}(Y_o - \bar{Y}_o)^2 \cdot \sum_{i=1}^{N}(Y_c - \bar{Y}_c)^2\right]^{1/2}} \qquad (7)$$

3  *SSE:* The SSE can be defined as:

$$SSE = \sum_{i=1}^{N}(Y_o - Y_c)^2 \qquad (8)$$

4  *MSE:* The MSE between observed and computed outputs can be defined as:

$$MSE = \frac{\sum_{i=1}^{N}(Y_c - Y_o)^2}{N} \qquad (9)$$

5  *RMSE:* The RMSE between observed and computed outputs can be defined as:

$$RMSE = \sqrt{\frac{\sum_{i=1}^{N}(Y_c - Y_o)^2}{N}} \qquad (10)$$

6  *NMSE:* The NMSE between observed and computed outputs can be defined as:

$$NMSE = \frac{\frac{1}{N}\sum_{i=1}^{N}(Y_c - Y_o)^2}{\sigma_{obs}^2} \qquad (11)$$

7  η*:* The η can be defined as:

$$\eta = 1 - \frac{\sum_{i=1}^{N}(Y_c - Y_o)^2}{\sum_{i=1}^{N}(Y_o - \bar{Y}_o)^2} \qquad (12)$$

8  *MAE:* MAE can be defined as follows:

$$MAE = 1 - \frac{\sum_{i=1}^{N}|Y_c - Y_o|}{\sum_{i=1}^{N}|Y_o - \bar{Y}_o|} \qquad (13)$$

where $N$ represents the number of feature vectors prepared, $Y_o$ and $Y_c$ denote the observed and the simulated values of dependent variable respectively, $\bar{Y}_o$ and $\sigma^2_{obs}$ are the mean and the standard deviation of the observed dependent variable respectively.

## 4  Detection of attacks

Here, we will discuss propose detection system that is part of access router or can belong to separate unit that interact with access router to detect attack traffic. Entropy-based DDoS scheme (Shannon, 2001) is used to construct profile of the traffic normally seen in the network, and identify anomalies whenever traffic goes out of profile. A metric that captures the degree of dispersal or concentration of a distribution is sample entropy. Sample entropy $H(X)$ is

$$H(X) = -\sum_{i=1}^{N} p_i \log_2(p_i) \quad (14)$$

where $p_i$ is $n_i/S$. Here, $n_i$ represent total number of bytes arrivals for a flow $i$ in $\{t - \Delta, t\}$ and $S = \sum_{i=1}^{N} n_i$, $i = 1, 2 \ldots N$. The value of sample entropy lies in the range $0\text{-}\log_2 N$.

To detect the attack, the value of $H_c(X)$ is calculated in time window $\Delta$ continuously; whenever there is appreciable deviation from $X_n(X)$, various types of DDoS attacks are detected. $H_c(X)$ and $X_n(X)$ gives entropy at the time of detection of attack and entropy value for normal profile respectively.

## 5  Experimental setup and performance analysis

In this section, we evaluate our proposed scheme using simulations. The simulations are carried out using NS2 network simulator. We show that false positives and false negatives (or various error rates) triggered by our scheme are very less. This implies that profiles built are reasonably stable and are able to estimate strength of DDoS attack correctly.

Real-world internet type topologies generated using transit-stub model of GT-ITM topology generator is used to test our proposed scheme, where transit domains are treated as different internet service provider (ISP) network, i.e., autonomous system (AS). For simulations, we use ISP level topology, which contains four transit domains with each domain contain twelve transit nodes, i.e., transit routers. All the four transit domains have two peer links at transit nodes with adjacent transit domains. Remaining ten transit nodes are connected to ten stub domain, one stub domain per transit node. Stub domains are used to connect transit domains with customer domains, as each stub domain contains a customer domain with ten legitimate client machines. So total of 400 legitimate client machines are used to generate background traffic.

**Figure 1** A short scale simulation topology (see online version for colours)

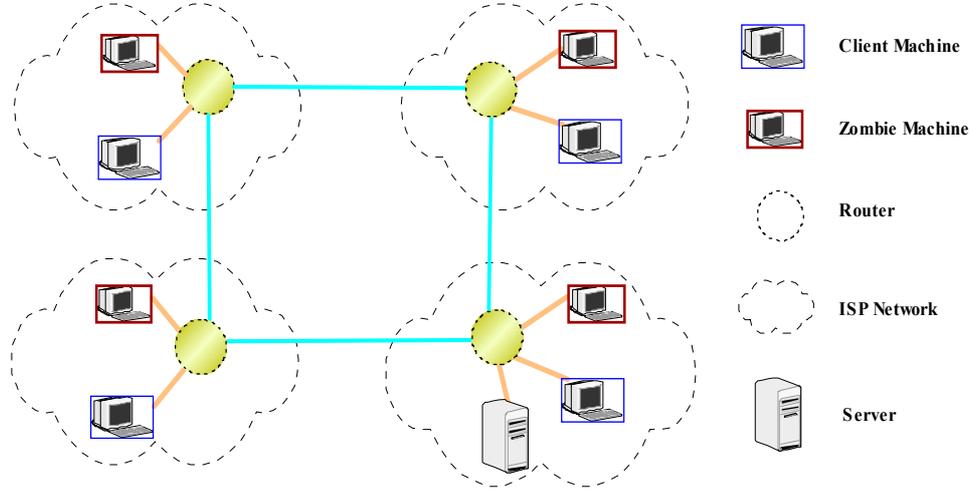

Total zombie machines are fixed to 100 to generate attack traffic. Transit domain four contains the server machine to be attacked by zombie machines. A short scale simulation topology is shown in Figure 1.

Currently, the majority of the DDoS attacks are flooding attack, so we will consider detection of a wide range of flooding DDoS attacks in this section. The legitimate clients are TCP agents that request files of size 1 Mbps with request inter-arrival times drawn from a Poisson distribution. The attackers are modelled by UDP agents. A UDP connection is used instead of a TCP one because in a practical attack flow, the attacker would normally never follow the basic rules of TCP, i.e., waiting for ACK packets before the next window of outstanding packets can be sent, etc. The attack traffic rate range between 10 Mbps and 100 Mbps; therefore, mean attack rate per zombie is varied from 0.1 Mbps to 1 Mbps. In our experiments, the monitoring time window was set to 200 ms, as the typical domestic internet RTT is around 100 ms and the average global internet RTT is 140 ms (Gibson, 2006). Total false positive alarms are minimum with high detection rate using this value of monitoring window. The simulations are repeated and different attack scenarios are compared by varying total attack strengths using fixed number of zombies.

## 6 Model development

In order to estimate strength of DDoS attack $\left(\hat{Y}\right)$ from deviation $(H_c - H_n)$ in entropy value, simulation experiments are done at varied attack strengths which are range between 10 Mbps and 100 Mbps and total number of zombies is fixed to 100; therefore, mean attack rate per zombie is varied from 0.1 Mbps to 1 Mbps.

Figure 2 shows entropy variation with varied attack strength, e.g., 10 Mbps to 100 Mbps while total number of zombies are fixed to 100. Table 1 represents deviation in entropy with actual strength of DDoS attack.

**Figure 2** Entropy variation with varied attack strength (see online version for colours)

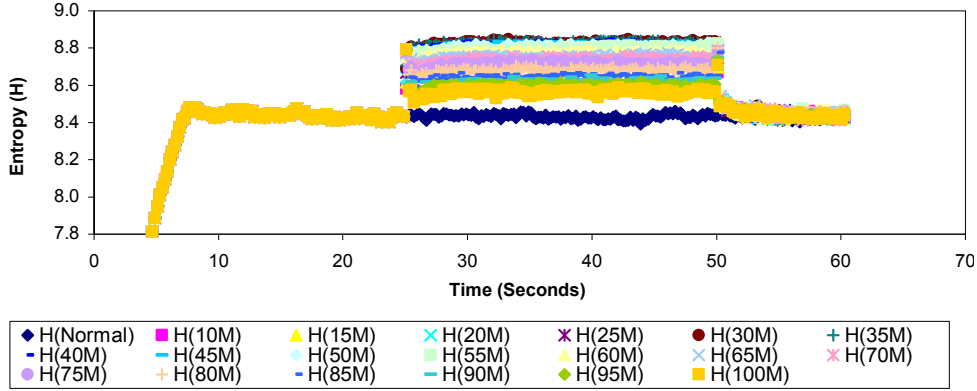

**Table 1** Deviation in entropy with actual strength of DDoS attack

| Actual strength of DDoS attack (Y) | Deviation in entropy (X) ($H_c - H_n$) |
|---|---|
| 10 M | 0.149 |
| 15 M | 0.169 |
| 20 M | 0.184 |
| 25 M | 0.192 |
| 30 M | 0.199 |
| 35 M | 0.197 |
| 40 M | 0.195 |
| 45 M | 0.195 |
| 50 M | 0.208 |
| 55 M | 0.212 |
| 60 M | 0.233 |
| 65 M | 0.241 |
| 70 M | 0.244 |
| 75 M | 0.253 |
| 80 M | 0.279 |
| 85 M | 0.280 |
| 90 M | 0.299 |
| 95 M | 0.296 |
| 100 M | 0.319 |

Various regression models, i.e., linear, polynomial, logarithmic, power and exponential are used to estimate strength of DDoS attack from deviation ($H_c - H_n$) in entropy value. Regression models are developed using strength of DDoS attack (*Y*) and deviation ($H_c - H_n$) in entropy value as discussed in Table 1 to fit the regression equation. Figures 3 to 7 show the regression equation and $R^2$ for each regression models as discussed in Section 2.

**Figure 3** Regression equation and $R^2$ for linear regression-based model M1 (see online version for colours)

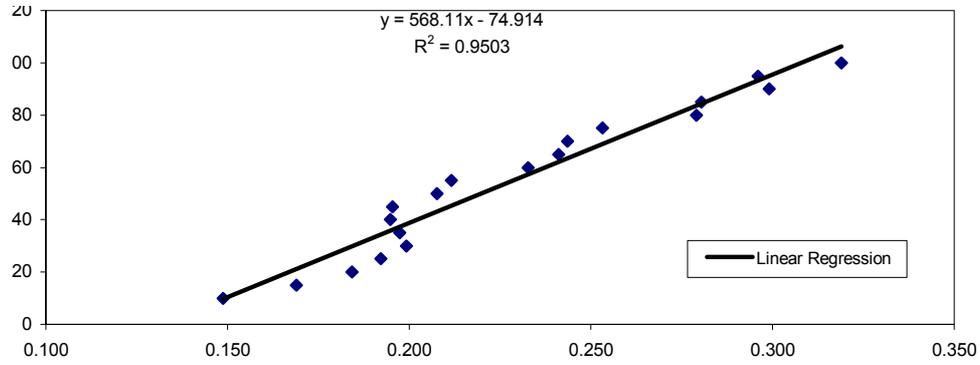

**Figure 4** Regression equation and $R^2$ for polynomial regression-based model M2 (see online version for colours)

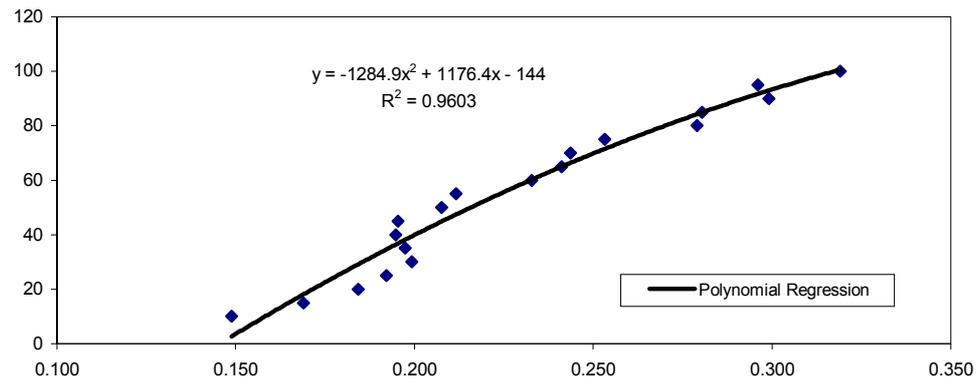

**Figure 5** Regression equation and $R^2$ for logarithmic regression-based model M3 (see online version for colours)

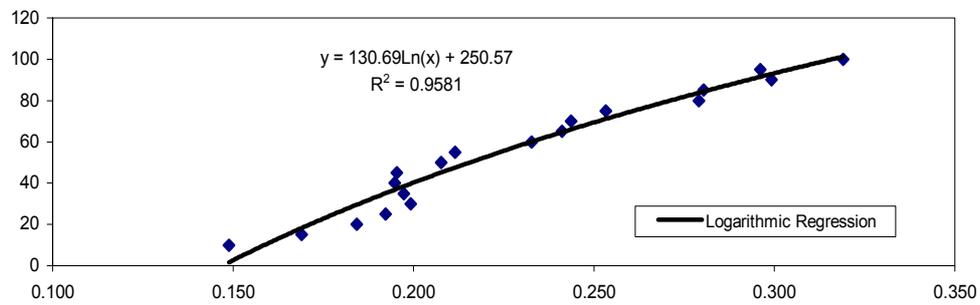

**Figure 6**   Regression equation and $R^2$ for power regression-based model M4 (see online version for colours)

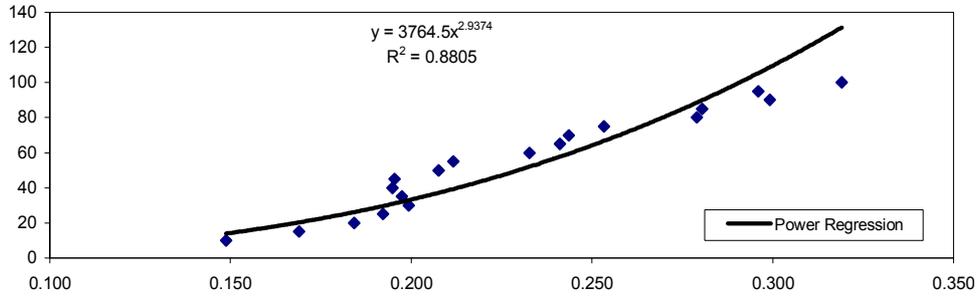

**Figure 7**   Regression equation and $R^2$ for exponential regression-based model M5 (see online version for colours)

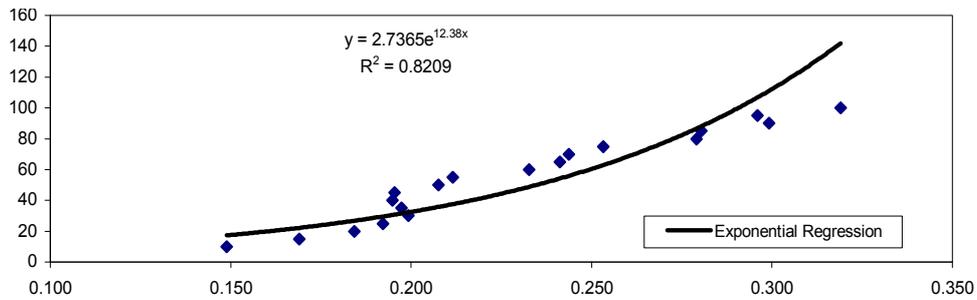

## 7   Result and discussion

We have developed models M1 to M5 as discussed in Section 6. Various performance measures are used to check the accuracy of these models. Constants $\beta_0$, $\beta_1$ for various regression equations are network environment dependent. Strength of DDoS attack can be computed and compared with actual strength of DDoS attack using various regression models. Figure 8 shows comparison between actual strength of DDoS attack and predicted strength of DDoS attack using various regression models M1 to M5 (viz. M1 to M5).

   To represent false positive and false negative, we plot residual error for various regression models. Positive cycle of residual error curve represents false positive, while negative cycle represents false negative. Figure 9 depicted summary of residual error in various regression models. Table 2 shows summery of various performance measures.

   It can be inferred from Figures 8 to 9 and Table 2 that polynomial regression-based model M2 performs better than other models. Values of $R^2$, CC, SSE, MSE, RMSE, NMSE, $\eta$, MAE are 0.96, 0.98, 566.31, 29.81, 5.46, 1.06, 0.96 and 0.81 respectively. Hence, strength of DDoS attack estimated by this model is closest to the observed strength of DDoS attack.

**Figure 8** Comparison between actual strength of DDoS attack and predicted strength of DDoS attack using various regression models M1 to M5 (see online version for colours)

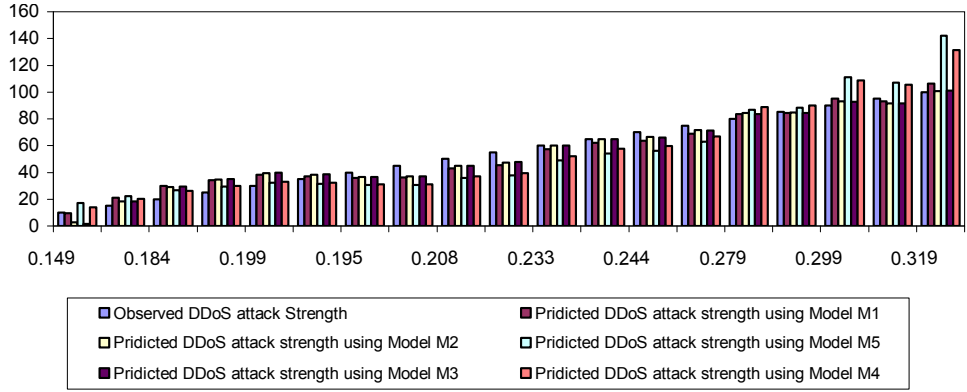

**Figure 9** Summary of residual error in various regression models (see online version for colours)

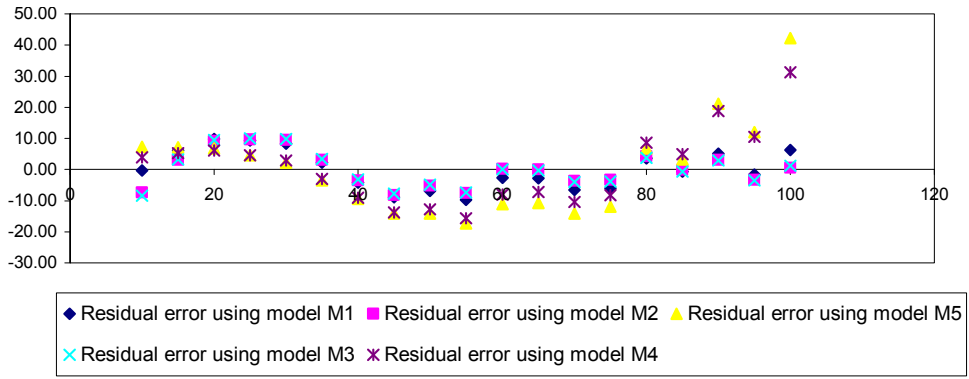

**Table 2** Summary of various performance measures

|  | Linear | Polynomial | Logarithmic | Power | Exponential |
|---|---|---|---|---|---|
| $R^2$ | 0.95 | 0.96 | 0.96 | 0.89 | 0.84 |
| CC | 0.97 | 0.98 | 0.98 | 0.94 | 0.92 |
| SSE | 708.13 | 566.31 | 596.96 | 2,643.90 | 3,995.70 |
| MSE | 37.27 | 29.81 | 31.42 | 139.15 | 210.30 |
| RMSE | 6.10 | 5.46 | 5.61 | 11.80 | 14.50 |
| NMSE | 1.32 | 1.06 | 1.12 | 4.95 | 7.47 |
| η | 0.95 | 0.96 | 0.96 | 0.81 | 0.72 |
| MAE | 0.78 | 0.81 | 0.80 | 0.59 | 0.51 |

## 8 Conclusions and future work

This paper investigates suitability of various regression models to predict strength of DDoS attack from deviation $(H_c(X) - X_n(X))$ in sample entropy. In order to predict number of zombies, several models are developed using various regression techniques. For each regression model, we have calculated various statistical performance measures, i.e., $R^2$, CC, SSE, MSE, RMSE, NMSE, $\eta$, MAE and residual error. Based on the statistical measures polynomial regression-based model perform better than any other model explored in this study. For polynomial regression-based model, values of $R^2$, CC, SSE, MSE, RMSE, NMSE, $\eta$, MAE are 0.96, 0.98, 566.31, 29.81, 5.46, 1.06, 0.96 and 0.81 respectively. Therefore, predicted strength of DDoS attack using polynomial regression is close to observe/actual strength of DDoS attack. However, simulation results are promising as we are able to estimate strength of DDoS attack using various regression models efficiently with very less error rate but experimental using a real time test bed can strongly validate our results.

## References

Anderson, T.W. (1962) 'The choice of the degree of a polynomial regression as a multiple decision problem', *The Annals of Mathematical Statistics*, Vol. 33, No. 1, pp.255–265.

Aneuryn-Evans, G. and Deaton, A. (1980) 'Testing linear versus logarithmic regression models', *The Review of Economic Studies*, Vol. 47, No. 1, pp.275–291.

Baskerville, G.L. (1972) 'Use of logarithmic regression in the estimation of plant biomass', *Canadian Journal of Forest Research*, NRC Research Press.

Dette, H., Lopez, I.M., Rodriguez, I.M.O. and Pepelyshev, A. (2006) 'Maximin efficient design of experiment for exponential regression models', *Journal of Statistical Planning and Inference*, Vol. 136, No. 12, pp.4397–4418.

El-Shaarawi, A. and Viveros-Aguilera, R. (2006) 'Logarithmic regression', *Encyclopedia of Environmetrics*, doi:10.1002/9780470057339.val015.

Freedman, D.A. (2005) *Statistical Models: Theory and Practice*, Cambridge University Press.

Gibson, B. (2006) *TCP Limitations on File Transfer Performance Hamper the Global Internet*, White paper, available at http://www.niwotnetworks.com/gbx/TCPLimitsFastFileTransfer.htm.

Gowariker, V., Thapliyal, V., Kulshrestha, S.M., Mandal, G.S. et al. (1991) 'A power regression model for long range forecast of southwest monsoon rainfall over India', *Mausam*, Vol. 42, No. 2, pp.125–130.

Gupta, B.B., Misra, M. and Joshi, R.C. (2008) 'An ISP level solution to combat DDoS attacks using combined statistical based approach', *International Journal of Information Assurance and Security (JIAS)*, Vol. 3, No. 2, pp.102–110.

Gupta, B.B., Misra, M. and Joshi, R.C. (2009) 'Defending against distributed denial of service attacks: issues and challenges', *Information Security Journal: A Global Perspective*, Vol. 18, No. 5, pp.224-247.

Jammalamadaka, S.R. (2003) 'Introduction to linear regression analysis', *The American Statistician*, Vol. 57, No. 1, pp.67–67.

Lindley, D.V. (1987) 'Regression and correlation analysis', *New Palgrave: A Dictionary of Economics*, Vol. 4, pp.120–123.

Melas, V.B. (1978) 'Optimal designs for exponential regression', *Statistics*, Vol. 9, No. 1, pp.45–59.

Montgomery, D.C., Peck, E.A. and Vining, G.G. (2001) *Introduction to Linear Regression Analysis*, 3rd ed., 641 pages, ISBN:0-471-31565-6, Wiley, New York.

Moore, D., Shannon, C., Brown, D.J., Voelker, G. and Savage, S. (2006) 'Inferring internet denial-of-service activity', *ACM Transactions on Computer Systems*, Vol. 24, No. 2, pp.115–139.

NS Documentation (2010) available at http://www.isi.edu/nsnam/ns.

Shannon, C.E. (2001) 'A mathematical theory of communication', *ACM SIGMOBILE Mobile Computing and Communication Review*, Vol. 5, pp.3–55.

Stigler, S.M. (1971) 'Optimal experimental design for polynomial regression', *Journal of American Statistical Association*, Vol. 66, No. 334, pp.311–318.